\documentclass[a4paper,fleqn]{cas-sc}

\usepackage[authoryear]{natbib}
\usepackage{setspace}
\usepackage{lineno}
\usepackage{comment}

\def\tsc#1{\csdef{#1}{\textsc{\lowercase{#1}}\xspace}}
\tsc{WGM}
\tsc{QE}
\tsc{EP}
\tsc{PMS}
\tsc{BEC}
\tsc{DE}

\begin{document}
\let\WriteBookmarks\relax
\def\floatpagepagefraction{1}
\def\textpagefraction{.001}
\shorttitle{Delayed-feedback oscillators replicate the dynamics of multiplex networks}
\shortauthors{A. Zakharova and V. Semenov}

\title [mode = title]{Delayed-feedback oscillators replicate the dynamics of multiplex networks: wavefront propagation and stochastic resonance}                      

\author[1]{Anna Zakharova \corref{cor1}}[orcid=0000-0002-1499-6043]
\address[1]{Institut f\"{u}r Theoretische Physik, Technische Universit\"{a}t Berlin, Hardenbergstra{\ss}e 36, 10623 Berlin, Germany}

\author[2]{Vladimir V. Semenov \corref{cor2}}[orcid=0000-0002-4534-8065]
\address[2]{Institute of Physics, Saratov State University, 83 Astrakhanskaya str., 410012 Saratov, Russia}

\begin{abstract}
The widespread development and use of neural networks have significantly enriched a wide range of computer algorithms and promise higher speed at lower cost. However, the imitation of neural networks by means of modern computing substrates is highly inefficient, whereas physical realization of large scale networks remains challenging. Fortunately, delayed-feedback oscillators, being much easier to realize experimentally, represent promising candidates for the empirical implementation of neural networks and next generation computing architectures. In the current research, we demonstrate that coupled bistable delayed-feedback oscillators emulate a multilayer network, where one single-layer network is connected to another single-layer network through coupling between replica nodes, i.e. the multiplex network. We show that all the aspects of the multiplexing impact on wavefront propagation and stochastic resonance identified in multilayer networks of bistable oscillators are entirely reproduced in the dynamics of time-delay oscillators. In particular, varying the coupling strength allows suppressing and enhancing the effect of stochastic resonance, as well as controlling the speed and direction of both deterministic and stochastic wavefront propagation. All the considered effects are studied in numerical simulations and confirmed in physical experiments, showing an excellent correspondence and disclosing thereby the robustness of the observed phenomena. 
\end{abstract}


\begin{keywords}
bistability \sep noise \sep wavefront propagation \sep stochastic resonance \sep delayed-feedback oscillator \sep multilayer network \sep multiplexing \sep numerical simulation \sep electronic experiment
\end{keywords}

\maketitle

\doublespacing

\section{Introduction}
\label{sec:Introduction}
The presence of sufficiently long delay in single oscillators can lead to the appearance of the complex high-dimensional dynamics and provide for the observation of oscillatory regimes originally revealed in spatially-extended systems or ensembles of coupled oscillators (\cite{yanchuk2017}). In such a case, spatio-temporal phenomena are tracked down in the purely temporal dynamics of delayed-feedback oscillators by applying a spatio-temporal representation, which considers the delay interval $[0:\tau]$ in analogy with the spatial coordinate  (\cite{arecchi1992,giacomelli1996}). Using this approach, the delay dynamics has been found to sustain stable chimera states (\cite{garbin2015,marconi2015,romeira2016,brunner2018,semenov2018,yanchuk2019,semenov2023}), travelling waves (\cite{klinshov2017}), coarsening (\cite{giacomelli2012,semenov2018}) and nucleation (\cite{zaks2013}). The capability of delay systems to reproduce the high-dimensional phenomena has revolutionized the hardware implementation of recurrent neural networks. In particular, oscillators forced by delayed feedback can efficiently be exploited for reservoir computing as a single-node reservoir (\cite{appeltant2011,martinenghi2012,larger2017,brunner2018-2,duport2012,huelser2022,koester2022}). Moreover, time-delay reservoirs extended to deep networks can be implemented physically based on the dynamics of coupled nonlinear delayed-feedback oscillators (\cite{penkovsky2019}) or a single system with multiple time-delayed feedback loops (\cite{stelzer2021}). In addition, one can apply singleton time-delay systems as a spin network (or spin glass) for solving problems with non-deterministic polynomial-time hardness (so-called NP-hard problems). Particularly, Ising machines based on delayed-feedback bistable systems have broad prospects for acceleration of the optimization problem computations (\cite{boehm2019}).

\begin{figure}[t]
\centering
\includegraphics[width=0.98\textwidth]{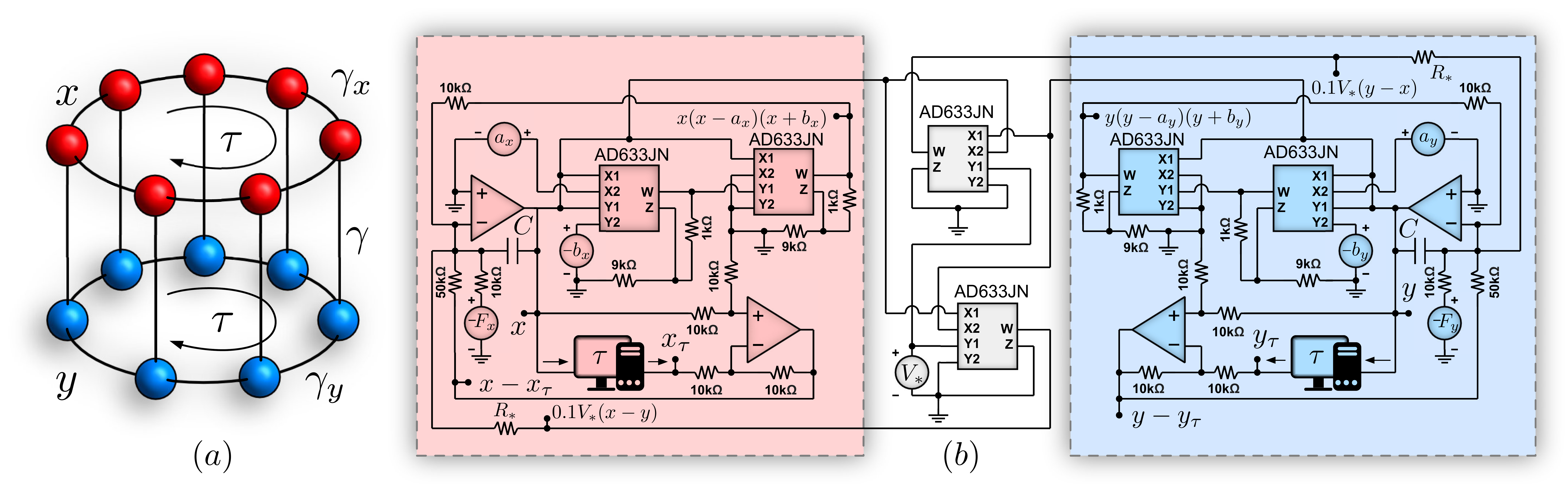} 
\caption{(a) Schematic illustration of the dynamics of coupled time delay systems; (b) Circuit diagram of the experimental setup (Eqs. (\ref{experimental_model})).}
\label{fig1}
\end{figure}  
In contrast to single delayed-feedback oscillators, coupled long-delay systems remain unexplored in the context of analogy to ensembles and networks (except of Ref. \cite{penkovsky2019} where the deep reservoir scheme comprises hierarchically coupled nonlinear delay oscillators). Meanwhile, the high-dimensional dynamics of coupled oscillators with long delay requires thorough analysis including research of the coupling role as well as its interpretation in the context of emulated network topology. In addition, it is not clear how accurately the dynamics of coupled delay oscillators replicate the collective behaviour in complex networks of coupled oscillators. In the current paper, we address this question and demonstrate that coupled delayed-feedback oscillators qualitatively reproduce the dynamics of multilayer networks where the interaction between the layers is realized through the coupling between replica nodes, i.e. multiplex networks (\cite{boccalletti2014,kivela2014}). Indeed, in the absence of coupling, delay systems are themselves equivalent to a ring of coupled oscillators with unidirectional coupling. In the presence of diffusive coupling, the oscillators affect each other at any time through the instantaneous values of dynamical variables. The described structure schematically illustrated in Fig.~\ref{fig1}~(a) resembles a two-layer multiplex network. 

To prove the statement about the similarity to multiplex networks, we show that the intrinsic peculiarities of such networks can be recognized in the dynamics of coupled time-delay oscillators. The distinguishable character of the multiplexing impact on the network dynamics has been
repeatedly reported and is responsible for various nontrivial phenomena and control schemes: intra-layer synchronization (\cite{gambuzza2015}), control of deviations from synchronization in presence of large perturbations (\cite{olmi2024}), reaching the inter-layer synchronization at lower inter-layer coupling strength (\cite{eser2021}) and multiplexing-noise-induced synchronization (\cite{rybalova2022}), control of chimera (\cite{zakharova2020}) and solitary (\cite{schuelen2021}) states and transitions between them (\cite{mikhailenko2019,rybalova2021}), control of coherence (\cite{masoliver2021,semenova2018}) and stochastic (\cite{semenov2022}) resonances and the wavefront propagation (\cite{semenov2023-2}), topological-asymmetry-induced stabilization of chaotic dynamics (so-called asymmetry-induced order) (\cite{medeiros2021}). Of particular interest are the phenomena occurring in a broad spectrum of bistable dynamical systems, namely the wavefront propagation (\cite{semenov2023-2}) and stochastic resonance (\cite{semenov2022}). Both of them are observed and can be efficiently controlled in multiplex networks.

The first effect, wavefront propagation, results from the presence of the asymmetry and can be controlled in a wide variety of dynamical systems including multiplex networks (\cite{semenov2023-2}) by varying the system parameters or by applying multiplicative noise (\cite{engel1985,loeber2014,loecher1998}). The second effect, stochastic resonance (\cite{gammaitoni1998,neiman1998,anishchenko1999}), is manifested through the noise-induced regularity of the stochastic dynamical system response to an input signal and can be enhanced or suppressed by varying the multiplexing strength (\cite{semenov2022}). 

\section{Model under study}
\label{sec:Model}
Both phenomena discussed in the current paper are explored on the example of a system of two coupled bistable delayed-feedback oscillators:
\begin{equation}
\label{numerical_model}
\begin{array}{l}
\dfrac{dx}{dt}=-x(x-a_x)(x+b_x)+\gamma_x(x_{\tau}-x)+\gamma(y-x)+F_x(t),\\
\dfrac{dy}{dt}=-y(y-a_y)(y+b_y)+\gamma_y(y_{\tau}-y)+\gamma(x-y)+F_y(t),\\
\end{array}
\end{equation}
where $\tau$ is the delay time, $x_{\tau}$ and $y_{\tau}$ are the values of variables $x$ and $y$, respectively, at the time moment $(t-\tau)$ with $\tau=1000$. The parameters $\gamma_{x}=\gamma_{y}=0.2$ characterize the delayed-feedback strength considered as an analog of the intra-layer coupling strength in the multiplex networks. The coupling strength $\gamma$ represents the strength of the inter-layer interaction in the multiplex network also called the strength of multiplexing (schematically illustrated in Fig.~\ref{fig1}~(a)). The external influences are given by $F_{x}$ and $F_{y}$. The parameters $a_{x},a_{y},b_{x},b_{y}>0$ define whether the oscillators nonlinearity in the bistable regime is symmetric ($a_x=b_x$ and $a_y=b_y$) or asymmetric ($a_x \neq b_x$ and $a_y \neq b_y$). 

Our investigations are performed by means of numerical simulations and electronic experiments. In more detail, we integrate stochastic differential equations (\ref{numerical_model}) numerically using the Heun method (\cite{mannella2002}) (see details in appendix \ref{app:num_sim}). For physical experiments, we have developed an experimental prototype being an electronic model of system (\ref{numerical_model}) implemented by principles of analog modelling (\cite{luchinsky1998}). The circuit diagram of the setup is shown in Fig.~\ref{fig1}~(b). The detailed description of the circuit is presented in appendix \ref{app:exp_setup}. The experimental setup is described by the equations 
\begin{equation}
\label{experimental_model}
\begin{array}{l}
RC\dfrac{dx}{dt}=-x(x-a_x)(x+b_x)+\gamma_x(x_{\tau}-x)+\gamma(y-x)+F_x(t),\\
RC\dfrac{dy}{dt}=-y(y-a_y)(y+b_y)+\gamma_y(y_{\tau}-y)+\gamma(x-y)+F_y(t),\\
\end{array}
\end{equation}
where $R=10k\Omega$, $C=10$ nF, $\gamma_x=\gamma_y=0.2$, $\gamma=0.1V_*R/R_*$, $\tau=27$ ms. Numerically and experimentally obtained time realizations $x(t)$ and $y(t)$ are mapped onto space-time ($\sigma,n$) by introducing $t=n\eta+\sigma$ with an integer time variable $n$, and a pseudo-space variable $\sigma \in [0,\eta]$, where $\eta=\tau+\varepsilon$ with a quantity $\varepsilon$, which is small compared to $\tau$ and results from a finite internal response time of the system (see the estimation of the resulting $\eta$ in appendix \ref{app:space_length}).

\section{Wavefront propagation}
\label{sec:wavefront_prop}

\begin{figure}[t]
\centering
\includegraphics[width=0.5\textwidth]{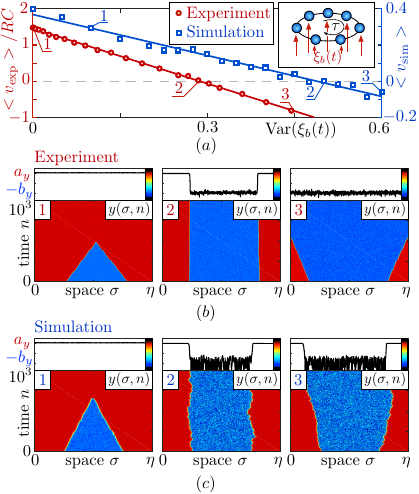} 
\caption{Stochastic wavefront propagation control in a single delayed-feedback oscillator $y$ (see Eqs. (\ref{numerical_model}) and (\ref{experimental_model}) by applying parametric noise $\xi_b(t)$: $b_y=0.45+\xi_b(t)$): (a) Dependence of the wavefront propagation speed on the stochastic influence's variance registered in numerical modelling and electronic experiments. Solid lines are obtained from a fit to the data using the least squares method: $v_{\text{exp}}=1.455-5.078$Var($\xi_b(t)$) (red solid line) and $v_{\text{sim}}=0.369-0.756$Var($\xi_b(t)$) (blue solid line). Space-time diagrams in panels (b) and (c) illustrate the stochastic dynamics in points 1-3 in numerical and experimental dependencies on panel (a). The upper insets in panels (b) and (c) show the oscillator's state at $n=10^3$. Other parameters: $a_y=0.5$, $\gamma_y=0.2$ ,$\gamma=0$, $\tau=1000$ (simulations) and $\tau=0.027$ ms (experiments), $F_{x}(t)\equiv 0$, $F_{y}(t)\equiv 0$.}
\label{fig2}
\end{figure}

Numerical simulations and experiments with analog circuit have shown that the studied delayed-feedback oscillators exhibit propagating fronts in the presence of asymmetry. Thus, the simplest way to control the front propagation speed is varying parameters $a_{x}$, $a_{y}$, $b_{x}$ and $b_{y}$. The second approach, stochastic control by applying multiplicative noise, is illustrated in Fig.~\ref{fig2} on an example of single oscillator $y$ in the presence of asymmetry and parametric Gaussian white noise (see the noise source description in  appendix \ref{app:noise}), and in the absence of external influence: $a_y=0.5$, $b_y=0.45+\xi_b(t)$, $F_{y}(t)\equiv 0$. To quantitatively describe the noise-controlled wavefront propagation, the wavefront propagation speed is introduced as the expansion speed of the state $y(t)=a_y$ (the red domain in Fig. \ref{fig2} (b),(c)) (see appendix \ref{app:space_length}). The dependence of the mean front propagation speed on the noise variance obtained in numerical simulations, $<v_{\text{sim}}(\text{Var}(\xi_b(t)))>$, and electronic experiments, $<v_{\text{exp}}(\text{Var}(\xi_b(t)))>$, clearly indicates the option for noise-sustained stabilization and reversal of the wavefront propagation [Fig.~\ref{fig2}~(a)]. In the presence of weak noise the dynamics is almost the same as in the deterministic case where one observes expansion of the domain corresponding to the state $y(t) = a_y=0.5$ [Fig. \ref{fig2}~(b),(c), panels 1]. For an appropriate noise intensity, such propagation can be stopped (stabilized) by parametric noise [Fig. \ref{fig2}~(b),(c), panels 2] exactly as in spatially-extended systems (\cite{engel1985}) and ensembles (\cite{semenov2023-2}) and then made inverse [Fig. \ref{fig2}~(b),(c), panels 3] for further increasing noise intensity. The opposite effect is achieved if instead modulating parameter $b_y$, the parametric noise is introduced for the parameter $a_y$ in the form: $a_y = 0.5 + \xi_a(t)$. In such a case, increasing noise intensity speeds up the front propagation and the state $y = a_y$ invades the available space faster than in the deterministic case. 

The results in Fig.~\ref{fig2}~(a) indicate the linear character of the dependence of the wavefront propagation speed on the noise signal variance. To illustrate this fact, we used curve-fitting based on the least squares method (see the solid lines in Fig.~\ref{fig2}~(a)). Slight deviations from the linear functional relationship observed in numerical model result from more pronounced fluctuations of the wavefronts as compared to experimentally registered ones (compare space-time plots in panels 2 of Fig.~\ref{fig2}~(b),(c)). To minimize the impact of this effect, we use the mean front propagation speed for the stochastic dynamics description.

\begin{figure}[t!]
\centering
\includegraphics[width=0.55\textwidth]{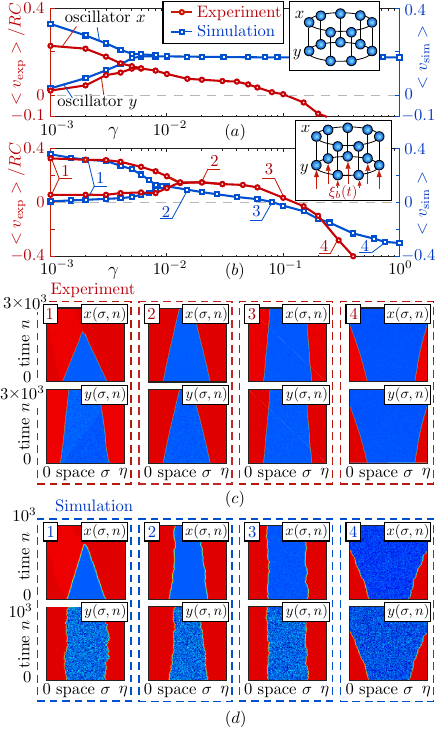} 
\caption{Control of deterministic and stochastic wavefront propagation by varying the coupling $\gamma$. Panel (a) illustrates the interaction of symmetric and asymmetric oscillators ($a_x=a_y=0.5$, $b_x=0.45$, $b_y=0.5$), whereas panel (b) corresponds to interaction of two asymmetric oscillators, one of which contains a noise source to stabilize the front propagation as in panels 2 of Fig.~\ref{fig2}~(b),(c) ($a_x=a_y=0.5$, $b_x=0.45$, $b_y=0.45+\xi_b(t)$). Space-time diagrams in panels (c) and (d) illustrate dynamics in points 1-4 in numerical and experimental dependencies on panel (b). Other parameters are $\gamma_x=\gamma_y=0.2$, $\tau=1000$ (simulations) and $\tau=0.027$ ms (experiments). External driving is absent, $F_{x}(t)\equiv 0$ and $F_{y}(t)\equiv 0$}
\label{fig3}
\end{figure}  

It is important to note that the stochastic control of wavefront propagation in single time-delayed feedback oscillators in itself is a novel phenomenon. The fact that multiplicative noise has exactly the same impact on wavefront propagation in bistable media, ensembles and time-delay oscillators with long delay emphasizes the similarity between these classes of high-dimensional dynamical systems. The multiplicative-noise-based control scheme can be applied in addition to deterministic approaches based on tuning the system parameters or applying external periodic signals (one of such methods is successfully applied in delayed optical system in Ref. \cite{marino2014}).

Next, the dynamics of non-identical coupled oscillators for varying coupling strength $\gamma$ is analysed. In the deterministic case, the parameters are chosen such that oscillator $x$ is asymmetric ($a_x=0.5$, $b_x=0.45$), while the second oscillator, $y$, is symmetric ($a_y=b_y=0.5$). In such a case, the front propagation speeds in oscillators $x$ and $y$ approach each other and tend to identical non-zero value when the coupling strength is increased [Fig.~\ref{fig3}~(a)]. Once the front propagation speeds have become identical, they do not vary for further increasing $\gamma$ in numerical experiments. In contrast, experimental setup demonstrates variable front propagation speed with increasing $\gamma$ after the common regime is achieved. This is due to the factors inevitably present in experimental electronic setups. The experimental setup is sensitive to inaccuracies which influence the system's symmetry and hence the wavefront propagation speed. All the circuit active elements, such as operation amplifiers and analog multipliers, have non-zero output offsets. For this reason, it is impossible to realize an absolutely symmetric deterministic oscillator characterised by zero wavefront propagation speed. Moreover, the impact of offsets becomes stronger when increasing $\gamma$, especially when the increase of $\gamma$ is realized by decreasing the resistors $R_{*}$ (see Fig.~\ref{fig1}). 

In the same way, increasing coupling strength reduces the stochastic dynamics to a common regime. The front propagation speed in the interacting oscillators reaches the same value when both oscillators are asymmetric and the wavefront propagation in oscillator $y$ is stabilised by noise as in panels 2 of Fig.~\ref{fig2}~(b),(c) ($a_x=a_y=0.5$ and $b_x=0.45$, $b_y=0.45+\xi_b$). In contrast to the deterministic oscillator interaction, the front propagation speeds continue to vary with increasing $\gamma$ after the common regime is reached [Fig.~\ref{fig3}~(b)-(d)]. This provides for stabilization and reversal of the wavefront propagation in interacting oscillators by varying $\gamma$ both in numerical and physical experiments.

\section{Stochastic resonance}
\label{sec:stoch_resonence}
\begin{figure}[t!]
\centering
\includegraphics[width=0.5\textwidth]{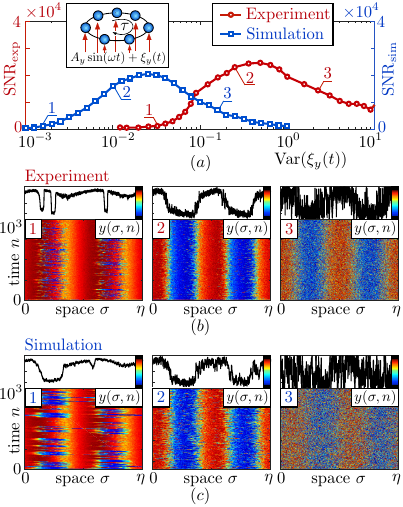} 
\caption{Stochastic resonance in a single delayed-feedback oscillator (oscillator $y$ in Eqs. (\ref{numerical_model}) and (\ref{experimental_model}) subject to periodic forcing and additive noise,  $F_y(t)=A\sin(\omega t)+\xi_y(t)$: (a) Signal-to-noise ratio as a function of the variance of noise. Space-time diagrams in panels (b) and (c) illustrate the stochastic dynamics in points 1-3 in numerical and experimental dependencies in panel (a). The upper insets in panels (b) and (c) show the oscillator's state at $n=10^3$. The system's parameters: $a_y=0.5$, $b_y=0.495$, $\gamma_y=0.2$, $\gamma=0$, $\tau=1000$ (simulations) and $\tau=0.027$ ms (experiments). External forcing parameters: $A=0.06$, $\omega=0.012$ (simulations) and $A=0.075$, $\omega=448.57$ (physical experiments).}
\label{fig4}
\end{figure}  

Now let us change the focus and consider single oscillator $y$ subject to external periodic forcing and additive Gaussian white noise, $F_y(t)=A\sin (\omega t)+\xi_y(t)$, to analyse the occurrence of stochastic resonance [Fig.~\ref{fig4}]. The oscillator's nonlinearity is slightly asymmetric $a_y=0.5$, $b_y=0.495$, which allows to avoid the impact of multistability. The period of the driving signal is close to half a delay time. The amplitude $A$ is smaller than the threshold value corresponding to the transformation of the dynamics into large-amplitude periodic jumps between two states. Under these conditions, the weak-noise-induced transitions between basins of attraction of two stable steady states $y^*_1=a_y$ and $y^*_2=b_y$ are infrequent (see panels 1 in Fig. \ref{fig4} (b),(c)). Most of the time the system oscillates near the steady state $y^*_1=a_y$. Noise of larger intensity leads to more and more frequent transitions between the basins of attraction. For an appropriate noise intensity, the transitions are the most regular and the corresponding space-time plots clearly indicate the periodicity in the dynamics of the non-autonomous oscillator [Fig.~\ref{fig4}~(b),(c), panels 2]. Increasing noise intensity even further leads to the descrease of the regularity of the system's response [panels 3 in Fig. \ref{fig4} (b),(c)]. To quantitatively describe the observed noise-induced dynamics, we introduce the signal-to-noise ratio (SNR) (see appendix \ref{app:snr}). The consideration of the SNR as a function of the noise variance in numerical and electronic experiments allows to obtain a non-monotonic curve being a signature of stochastic resonance: there exists an appropriate noise intensity corresponding to the maximal SNR [Fig.~\ref{fig4}~(a)].

\begin{figure}[t!]
\centering
\includegraphics[width=0.5\textwidth]{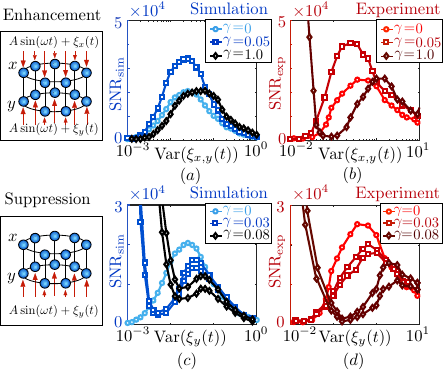} 
\caption{Enhancement (panels (a) and (b)) and suppression (panels (c) and (d)) of stochastic resonance by varying the coupling strength. Panels (a) and (b) illustrate the SNR for oscillations $y(t)$, whereas the SNRs corresponding to $\gamma=0.03$ and $\gamma=0.08$ in panels (c) and (d) are represented by two curves: the SNR characterizing the oscillations $x(t)$ (lower curves) and the oscillations $y(t)$ (the upper curves). External forces are: $F_x(t)=F_y(t)=A\sin(\omega t)+\xi_y(t)$ (panels (a) and (b)), $F_x(t)\equiv0$ and $F_y(t)=A\sin(\omega t)+\xi_y(t)$ (panels (c) and (d)). The parameter values are the same as in Fig. \ref{fig4} except of $\gamma$.}
\label{fig5}
\end{figure}  

If bistable delayed-feedback oscillators are coupled and forced by a common periodic signal and a source of additive Gaussian white noise, the effect of stochastic resonance can be enhanced. As demonstrated in Fig. \ref{fig5} (a),(b), one achieves higher signal-to-noise ratios in comparison to a single-system dynamics by varying the coupling strength. However, further increasing coupling returns the stochastic resonance manifestation to the initial, single-system form (except of horizontal shift). Note that the impact of coupling has the resonant character: there is an appropriate coupling strength corresponding to the most pronounced stochastic resonance.

In contrast, stochastic resonance can be suppressed by coupling, if either oscillator is not under direct action of noise and the periodic force [Fig. \ref{fig5} (c),(d)]. Since there is no interaction between oscillators $x$ and $y$ at $\gamma=0$, for this case Fig. \ref{fig5} (c),(d) displays the dependence of the SNR on the noise variance only for oscillator $y$. At non-zero $\gamma$, the curves SNR(Var($\xi_{y}(t)$)) are illustrated for both oscillators $x$ and $y$. Interestingly, we observe coupling-induced stochastic resonance in oscillator $x$, which receives periodic action and noise only through the coupling. However, increasing the coupling strength makes the effect of stochastic resonance in both oscillators less and less pronounced.

\section{Conclusion}
\label{sec:conclusion}
In conclusion, we find that the dynamics of coupled bistable delayed-feedback oscillators reproduces with high accuracy the effects in multiplex networks of bistable elements. In the context of this similarity, the delayed feedback strength plays a role of the coupling within the layer (intra-layer coupling), whereas the coupling between two delayed-feedback oscillators corresponds to the multiplex links between the layers (inter-layer coupling). To formulate this general conclusion, we demonstrate the revealed similarity on examples of different effects being typical for bistable dynamical systems: deterministic and stochastic wavefront propagation and stochastic resonance. Moreover, the reported numerical results are supported by a physical electronic experiment, which indicates the robustness of the observed phenomena. Obtained on examples of basic bistable delayed-feedback oscillators, the considered effects are expected to be similarly manifested in coupled bistable time delay oscillators of different nature.

Intriguingly, all the aspects of the multiplexing-based wavefront propagation control reported in Ref. \cite{semenov2023-2} are replicated by a numerical model of two oscillators with time delay. This is manifested through the reduction of the wavefront propagation to a common regime where the front propagation speeds tend to the same value. This value is fixed in the deterministic model, but varies in the stochastic oscillators with increasing the strength of multiplexing. This provides for stabilization and reversal of the front propagation in the interacting oscillators. It is important to note that due to the specific properties of the experimental setup, the common regime where the propagation speed is fixed with increasing coupling strength can not be achieved.
In addition, the intrinsic peculiarities of the multiplexing-based control of stochastic resonance established in Ref. \cite{semenov2022} have been revealed in a pair of coupled delayed-feedback oscillators. In more detail, coupling suppresses stochastic resonance, if the periodic forcing and noise are present in only one of the two oscillators. In contrast, increasing the coupling strength allows to enhance the stochastic resonance, if the periodic forcing and noise are present in all the interacting systems. In such a case, the impact of multiplexing has a resonant character.

The presented results have a wide spectrum of potential applications in the context of the delayed-feedback-oscillator-based physical implementations of machine learning algorithms initially realized by means of multilayer networks (\cite{penkovsky2019,stelzer2021}), especially the multiplex ones (\cite{amoroso2019}). In addition, a distinguishable application where the property of bistability and the symmetry control are of particular importance is the physical implementation of spin-networks as solvers of combinatorial optimization problems. This could be experimentally realized without classical computational machines based on the Von Neumann architecture (\cite{boehm2019,mohseni2022}). Another interesting field associated with control of delayed-feedback oscillators as spatially-extended systems by varying the coupling strength refers to the development of new schemes for controlling complex spatially-extended systems (for instance, see reaction-diffusion neural networks (\cite{song2023}) and dynamical surfaces (\cite{song2023-2})) and generalization of approaches for delayed-induced dynamics control (\cite{song2023-3}). 

It is important to note that in many cases single delayed-feedback oscillators are much easier to implement experimentally as compared to networks of coupled oscillators. Moreover, delayed-feedback oscillators are characterised by the property of scalability. Indeed, the quasi space length is fully determined by the delay time and can be extended without principal changes of the experimental setup (only a delay line is modified). In particular, delay time in the experimental setup depicted in Fig.~\ref{fig1}~(b) is realized electronically by using a personal computer and can be easily tuned by varying the buffer size defined in the Labview program (see appendix \ref{app:delay}). This solution is quite flexible and can be modified to avoid using personal computers. For this purpose, one can use programmable microcontrollers equipped by internal converters of digital and analog signals or complemented by external ones (for instance, such scheme was applied for delayed-feedback control of coherence resonance (\cite{semenov2015})). FPGA-based electronic systems also are appropriate candidates for implementations of electronic delay lines (\cite{boehm2019,boehm2022}). If one needs to create pure analog system, optical delay lines are well suited for this. For instance, optical fiber-based delay lines are used in the context of delay-based reservoir computing for spoken digit and speaker recognition and time series prediction (\cite{brunner2013,larger2017}) at data rates beyond 1 gigabyte per second. Choosing optical fibers of different length, materials and structure, one can create time delay in a wide range from pico- to microseconds. In summary, a wide diversity of currently known technical solutions for delay implementation offers significant opportunities for observation of the effects discussed in this paper. Optical fiber-based delayed-feedback is more preferable for high-frequency oscillators and applications. If it is principally important to obtain the delayed signal with minimal inaccuracies and the characteristic response time of delayed-feedback oscillator exceeds 70 $\mu$s, the PC-based delay line described in appendix \ref{app:delay} is a perfect solution. 

Despite the mentioned advantages, one must be aware of limitations when implementing delayed-feedback oscillators and physical realizing machine learning algorithm adapted to such dynamical systems (for instance, see delay-oscillator-based reservoir computing). First of all, the delay line is realized digitally by using analog-to-digital and digital-to-analog converters (see appendix \ref{app:delay}) which causes the emergence of an additional kind of noise, the measurement noise. Such stochastic factor can noticeably affect the dynamics close to bifurcation points where dynamical systems are structurally unstable. Second, the effects described in the paper are significantly dependent on inaccuracies and offsets inevitably presenting in the physical model and influencing the oscillators' asymmetry (see appendix \ref{app:exp_setup}). That means one must realize electronic models of bistable delayed-feedback oscillators as precisely as possible to experimentally observe the  phenomena of wavefront propagation and stochastic resonance and to control them. Finally, the general aspect of the complex dynamics associated with delay must be noted: the similarity between the delayed feedback systems and networks of coupled oscillators does not mean the full correspondence and will not necessarily eliminate the appearance of additional effect such as delay-induced multistability, bifurcations and chaos. For this reason, applying oscillators with delay for replication of the network dynamics is usually accompanied by consideration of individual properties of the oscillatory dynamics such as analysis of basins of attraction and response to external signals (input data). Thus, further studies focused on applications of bistable oscillators with delay as multiplex networks are especially intriguing. 

\section*{Declaration of Competing Interest}
The authors declare that they have no known competing financial interests or personal relationships that could have appeared to influence the work reported in this paper.

\section*{Data Availability}
The data that support the findings of this study are available from the corresponding author upon reasonable request.

\section*{Acknowledgements}
Vladimir Semenov acknowledges support by the Russian Science Foundation (project No.  22-72-00038).

\section*{Appendices}
\appendix

\section{Details of numerical simulations}
\label{app:num_sim}
Heun's method (also called improved or modified Euler's method) was applied to model system (\ref{numerical_model}) for further time series analysis. Since the equations contain the delay term, one needs to specify the initial conditions in the time range [$-\tau:0$) before the simulation starts.  To induce travelling fronts and to obtain elements of Fig.~\ref{fig2}~(a),(c) and Fig.~\ref{fig3}~(a),(b),(d) the following initial conditions were used: 
\begin{equation}
\label{experimental_model}
\begin{array}{l}
t \in \left[-\tau : -\dfrac{3}{4}\tau \right) \land \left[-\dfrac{1}{4}\tau : 0 \right): \quad x_0=a_x \quad \text{and} \quad y_0=a_y,\\
t \in \left[-\dfrac{3}{4}\tau : -\dfrac{1}{4}\tau \right): \quad x_0=b_x \quad \text{and} \quad y_0=b_y.
\end{array}
\end{equation}
To study the effect of stochastic resonance numerically (see Fig.~\ref{fig4}~(a),(c) and Fig. \ref{fig5} (a),(c)), the uniformly distributed initial conditions were used $x_0\in(0.25:0.75)$ and $y_0\in(0.25:0.75)$. 

To study both wavefront propagation and stochastic resonance, system (\ref{numerical_model}) was integrated with the time step $h=0.01$.

\section{Numerical and experimental noise sources and their characteristics}
\label{app:noise}
Numerical simulations involve a model of white Gaussian noise with intensity $D$: $\left<\xi_b(t)\right>=0$, $\left<\xi_b(t)\xi_b(t + \Delta t)\right> = 2D\delta(\Delta t)$. In this way, the variance of noise equals to $2D$. A source of noise used in physical experiments, Agilent 33250A, produces broadband Gaussian noise, whose spectral density was almost constant in a wide frequency range up to several tens of MHz. In this frequency range noise can be approximated by white Gaussian. 

\section{Full description of the experimental setup}
\label{app:exp_setup}
\begin{figure*} [t!]
  \begin{center}
    \includegraphics[width=15cm]{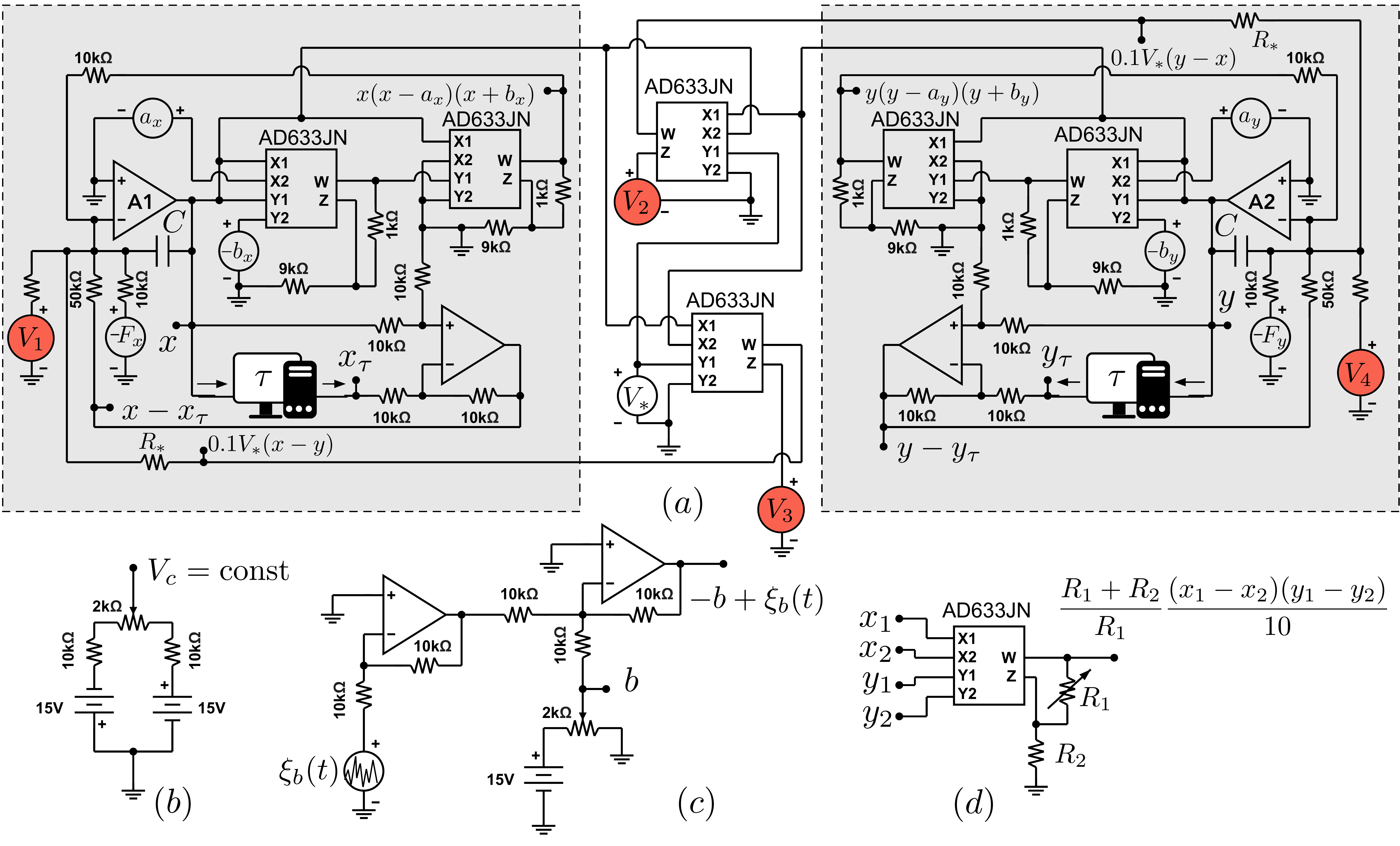}
  \end{center}
  \caption{Detailed circuit diagram of the experimental setup (panel (a)) and its particular elements (panels (b)-(d)). Operational amplifiers are TL072CP.} 
  \label{fig1_suppl}
\end{figure*}

Figure~1~(b) in the main manuscript illustrates the experimental setup circuit in the simplified form. A more detailed diagram is depicted in Fig. \ref{fig1_suppl} (a). It contains two integrators, A1 and A2, whose output voltages are taken as the dynamical variables, $x$ and $y$, respectively. The experimental facility includes a time-delay line, which was realized using personal computer complemented by an acquisition board (National Instruments NI-PCI 6251). The time delay value is constant, $\tau=27$ ms. Experimental setup equations (\ref{experimental_model}) include quantities $R=10k\Omega$ and $C=10$ nF being the resistance and the capacitance at the integrators A1 and A2. The resistance $R$ and the capacitance $C$ are responsible for system's response time: the higher are $R$ and $C$, the larger is the system's response time. The best resistor values are within the range 1-100 k$\Omega$, while the best capacitance values must be in the range 1-100 nF. Similar values for analog model circuit elements are recommended in review (\cite{luchinsky1998}) and are dictated by practical experience. If it is necessary to implement an electronic model characterised by low response time (high frequencies), then the used operational amplifiers must be chosen with taking into account the frequencies of operational amplifiers' input signals and the operational amplifiers' slew rate characteristic. For more detail, one can find all the necessary information regarding OA-based analog integrators, inverting amplifiers, non-inverting amplifiers, their elements (resistors and capacitors) in review (\cite{luchinsky1998}). The parameters $\gamma_x$ and $\gamma_y$ in Eqs. (\ref{experimental_model}) are $\gamma_x=\gamma_y=R/50k\Omega=0.2$. The expression for coupling strength is $\gamma=0.1V_*\dfrac{R}{R_*}$, which means the simultaneous dependence on $V_*$ and $R_*$. This configuration allows to vary the parameter $\gamma$ in a wide range. The equations of the experimental setup can be transformed into the initial dimensionless model under consideration by using substitution $t=t/\tau_0$  ($\tau_0=RC=0.1$~ms is the circuit's time constant) and new dynamical variables $x/V_{0}$ and $y/V_{0}$, where $V_{0}$ is the unity voltage, $V_{0}=1$~V.

During the development of the experimental setup, the main problem we dealt with is the circuit's sensitivity to the individual non-zero output offset voltages of the integrated circuits being responsible for the coupling (two analog multipliers AD633JN in the middle of panel (a) in Fig. \ref{fig1_suppl}) and the signal integration process (operational amplifiers A1 and A2 in Fig. \ref{fig1_suppl} (a)). It is important to note that such offsets influence the oscillators' asymmetry. To minimize this effect and improve the setup, a sources of low DC-voltages (do not exceed 20mV) are included into the circuit (red sources $V_1$-$V_4$ in Fig. \ref{fig1_suppl} (a)). These sources are implemented using the potentiometer as in Fig. \ref{fig1_suppl} (b). The potentiometers being responsible for $V_1$ and $V_4$ are tuned such that the experimentally observed stable steady state regimes of single oscillators precisely coincide with expected values $a$ and $-b$ (two coexisting stable equilibria of the bistable model $\dot{u}=-u(u-a)(u+b)+\gamma_u(u_{\tau}-u)$ are $u_*=a$ and $u_*=-b$). Voltages $V_2$ and $V_3$ are the multipliers' offsets $Z$ and are adjusted such that the zero inputs (when all the multiplier's input pins are connected to ground) result in zero output voltage. 

When the experiments on the stochastic wavefront propagation control were carried out, the noise modulation of the parameter $b_y$ was implemented as in Fig. \ref{fig1_suppl} (c). The voltage divider allows to adjust DC-voltage $b$ with high accuracy. The resulting signal is the sum $-b+\xi(t)$. To realize the regime of wavefront propagation from the almost the same initial states, periodic square wave signals $F_x(t)$ and $F_y(t)$ were used for several seconds to provide starting from identical initial conditions. Then the external signals were switched off and the experimentally observed evolution was registered.

To precisely realise the oscillators' cubic nonlinearities, the scale factor of multipliers AD633JN is tuned using a variable resistor (see the datasheet for AD633) as in Fig. \ref{fig1_suppl} (d). It is recommended to accurately tune the scale factor to achieve precise  multiplication with the scale factor $0.1(R_1+R_2)/R_1$.

\section{Delay line implementation}
\label{app:delay}

\begin{figure} [t!]
  \begin{center}
    \includegraphics[width=12cm]{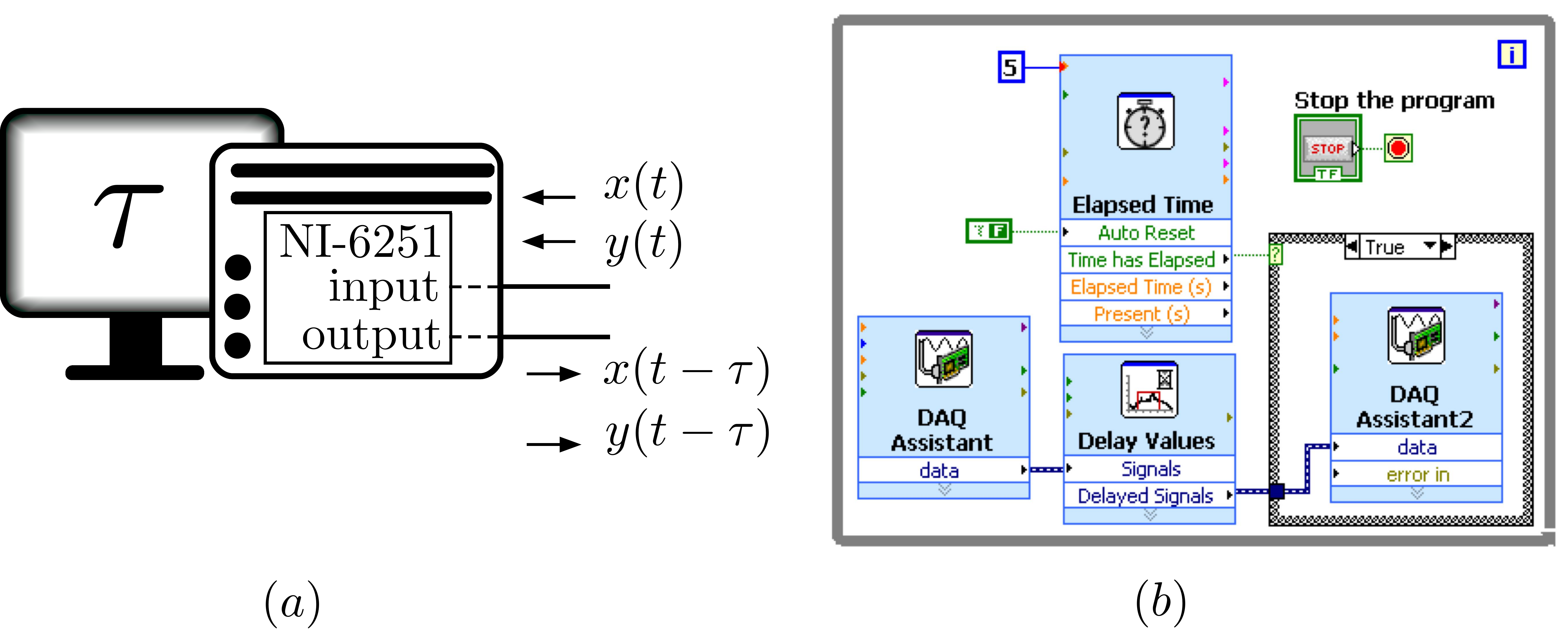}
  \
  \end{center}
  \caption{(a) Schematic representation of a PC-based delay line; (b) Block diagram of the LabVIEW program developed for the PC-based delay line implementation. The program is illustrated for case 'True' corresponding to obeying the condition. If the condition is not obeyed (case 'False'), the condition block is empty and the output signals are not generated.} 
  \label{fig_suppl_delay}
\end{figure}

The approach used for the development of the delay line involves a personal computer as a part of the experimental setup and provides for simultaneous generation of two delayed signals for implementing a system of two delayed-feedback oscillators with identical delay times. The used computer is equipped by a board NI PCI-6251 manufactured by National Instruments (schematically illustrated in Fig.~\ref{fig_suppl_delay}~(a)). The board contains analog-to-digital and digital-to-analog converters to acquire the input analog signals $x(t)$, $y(t)$ and to convert them into a digital form for further processing by PC. After the delayed signals $x(t-\tau)$ and $y(t-\tau)$ are saved as arrays in computer's memory, they are generated as analog output signals. To realize the processes of data acquisition, transformation and output signal generation, the program on the platform LabVIEW was developed. The block diagram of the program (see Fig.~\ref{fig_suppl_delay}~(b)) contains two DAQ-assistant instruments. The first block 'DAQ Assistant' is used to configure the input data acquisition (it contains a set of parameters such as a regime of acquisition, a list of used input channels, a sampling rate, an amplitude range, etc.). The second block 'DAQ Assistant 2' is responsible for analog output signal generation and allows to set the corresponding parameters. It is important to note that the output signal generation begins 5 seconds after the program is launched to provide for the initialization of all the processes. To realize it, the program contains a block acting as the 'if' operator. Panel (b) in Fig.~\ref{fig_suppl_delay} illustrates the case of obeying the condition of 'elapsing 5 seconds' (this block is empty for not obeying the condition). Inside the program, the delayed signal is produced by the block 'Delay Values' situated between the DAQ-assistant blocks. It contains a controllable parameter being responsible for delay time. The minimal and maximal possible delay times depend on the personal computer characteristics. 

\section{Estimation of the quasi-space length and introduction of the wavefront propagation speed}
\label{app:space_length}
\begin{figure} [t!]
  \begin{center}
    \includegraphics[width=14cm]{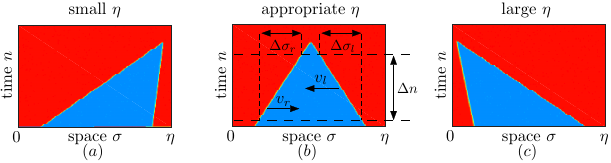}
  \
  \end{center}
  \caption{Methodology for choosing the space length $\eta$ and the front propagation speed $v$.} 
  \label{fig2_suppl}
\end{figure}

A unique value of the quasi-space length $\eta=\tau+\varepsilon$ is chosen in numerical and physical experiments according to the same principle: to achieve vertically oriented space-time diagrams $x(\sigma,n)$ and $y(\sigma,n)$. If the external forcing is absent, $F_{x}\equiv0$ and $F_{y}\equiv0$, this corresponds to the vertically symmetric diagrams such that the propagation to the left and to the right is identical as in the classical case of wavefront propagation speed in media and ensembles. Choosing an appropriate value of $\eta=\tau+\varepsilon$ is illustrated in Fig.~\ref{fig2_suppl}. In particular, the depicted in Fig.~\ref{fig2_suppl}~(b) space-time plot corresponds to the best value of $\eta$ and contains two identical fronts propagating to the left and to the right with identical speed. 

After the appropriate value of $\eta$ is found, the wavefront propagation speed is calculated (see Fig.~\ref{fig2_suppl}~(b)). In the current paper, we characterise the wavefront propagation by the expansion speed of the state $a$ (the red domain). For this purpose, the propagation speed is introduced as the averaged speed involving both speeds to the left, $v_l$ and to the right, $v_r$, as being:  $v=(v_{l}+v_{r})/2$. The speeds $v_{l}$ and $v_{r}$ are considered as the ratios $v_{l}=\Delta\sigma_{l}/\Delta n_{l}$ and $v_{r}=\Delta\sigma_{r}/\Delta n_{r}$. Such approach for calculation $v$ allows to avoid the dependence on choosing the parameter $\eta$ (the slope in space-time diagrams influences both $v_{l,r}$ such that the increase of resulting $v_{l}$ is accompanied by the same decreasing $v_{r}$, and vice versa).

\begin{figure} [t!]
  \begin{center}
    \includegraphics[width=9.5cm]{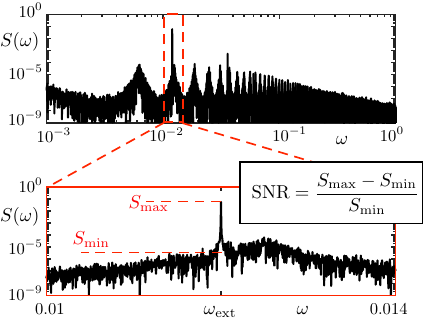}
  \end{center}
  \caption{Methodology for calculation of the signal-to-noise ratio using the power spectrum $S(\omega)$.} 
  \label{fig3_suppl}
\end{figure}

When the delayed-feedback oscillators are forced by external periodic forcing $A\sin(\omega t)$ and the period of external signal is close to $\tau/k$ (where $k\in\mathbb{N}$), the quasi-space length becomes entrained and equal to $\eta=2\pi/k\omega$ for the large enough  force's amplitude $A$. In such a case, the dynamics of the systems becomes locked to the frequency of the external driving.
In summary, all the values of $\eta$ used in the main manuscript for the visualization of the dynamics in the quasi-space are listed below:
\begin{itemize}
\item \textbf{Figure 2 (b)}: $\eta=0.0272282$ (panel (1)), $\eta=0.027225$ (panel (2)), $\eta=0.027226$ (panel (3))
\item \textbf{Figure 2 (c)}: $\eta=1005.95$ (panel (1)), $\eta=1005.55$ (panel (2)), $\eta=1005.45$ (panel (3))
\item \textbf{Figure 3 (c)}: $\eta_x=0.027213$, $\eta_y=0.02721$ (panel (1)), $\eta_x=\eta_y=0.0272116$ (panel (2)), $\eta_x=\eta_y=0.027226$ (panel (3)), $\eta_x=\eta_y=0.027229$ (panel (4)) 
\item \textbf{Figure 3 (d)}: $\eta_x=1005.53$, $\eta_y=1005.93$ (panel (1)), $\eta_x=\eta_y=1005.7$ (panel (2)), $\eta_x=\eta_y=1005.78$ (panel (3)), $\eta_x=\eta_y=1005.92$ (panel (4))
\item \textbf{Figure 4 (b)}: $\eta=0.028$ (panels (1)-(3))
\item \textbf{Figure 4 (c)}: $\eta=1047.2$ (panels (1)-(3))
\end{itemize}

\section{Calculation of the signal-to-noise ratio (SNR)}
\label{app:snr}
We use the method of calculation of the signal-to-noise ratio being a common measure for stochastic resonance. The typical power spectrum for both single and coupled delayed-feedback oscillators under stochastic driving is depicted in Fig.~\ref{fig3_suppl}. It includes the spectral peak $S_{\text{max}}$ at the frequency of external forcing, $\omega_{\text{ext}}$, which is also the main peak in the power spectrum. The power spectrum also has a minimum $S_{\text{min}}$ close to the spectral peak. In radiophysics the most common definition of the signal-to-noise ratio (SNR) is SNR$= P_{\text{S}}/P_{\text{N}}$, where $P_{\text{S}}$ is the power of the signal and $P_{\text{N}}$ is the noise power. The following formula of SNR corresponds to the harmonic external input signal in experiments: SNR$=H_{\text{s}}/H_{\text{n}}$, where  $H_{\text{s}}$ is the height of the spectral line above the background noise level in the power spectrum, and $H_{\text{n}}$ is the background noise level close to the resonance frequency $\omega_{\text{ext}}$, and thus in terms of the power spectrum SNR$=(S_{\text{max}}-S_{\text{min}})/S_{\text{min}}$. Technically, the value $S_{\text{min}}$ can be estimated as the mean value of the power spectrum $S(\omega)$ slightly to the left and to the right of the main spectral peak.


%

\end{document}